\documentclass[twocolumn,aps,prl,groupedaddress]{revtex4}
\usepackage{graphicx,wrapfig,amsmath,amssymb,enumitem,upgreek,bbm,mathtools,amsfonts,physics,multirow,color,caption,hyperref,xcolor,lineno}
\usepackage[normalem]{ulem}

\renewcommand{\vec}[1]{\oldbm{#1}}

\renewcommand{\vec}[1]{\boldsymbol{#1}}

\def\bb{{\vec b}}
\def\bk{{\vec K}}
\def\bk{{\vec k}}

\def\ba{{\vec a}}

\def\bK{{\vec K}}
\def\bq{{\vec q}}
\def\bQ{{\vec Q}}

\def\bG{{\vec G}}

\def\bn{{\vec n}}

\def\br{{\vec r}}

\def\bsigma{{\boldsymbol \sigma}}

\newcommand{\ov}[1]{\overline{#1}}

\def\tr{\mathop{\mathrm{tr}}}

\def\T{\mathcal{T}}

\def\D{\mathcal{D}}

\def\H{\mathcal{H}}

\def\tr{{\rm tr}}

\newcommand{\beq}{\begin{equation}}
\newcommand{\eeq}{\end{equation}}
\newcommand{\beqarray}{\begin{eqnarray}}
\newcommand{\eeqarray}{\end{eqnarray}}

\begin{document}

\title{
Family of Ideal Chern Flat bands with Arbitrary Chern Number in Chiral Twisted Graphene Multilayers
}

\author{Patrick J. Ledwith, Ashvin Vishwanath, Eslam Khalaf}
\affiliation{Department of Physics, Harvard University, Cambridge, Massachusetts 02138, USA}

\begin{abstract}
We consider a family of twisted graphene multilayers consisting of $n$-untwisted chirally stacked layers, e.g., AB, ABC, etc, with a single twist on top of $m$-untwisted {chirally stacked} layers. Upon neglecting both trigonal warping terms for the untwisted layers and the same sublattice hopping between all layers, the resulting models generalize several remarkable features of the chiral model of twisted bilayer graphene (CTBG).
In particular, they exhibit a set of magic angles which are identical to those of CTBG at which a pair of bands (i) are perfectly flat, (ii) have Chern numbers in the sublattice basis given by $\pm (n, -m)$ or $\pm (n + m - 1, -1)$ depending on the stacking chirality, and (iii) satisfy the trace condition, saturating an inequality between the quantum metric and the Berry curvature, and thus realizing ideal quantum geometry. These are the first higher Chern bands that satisfy (iii) beyond fine-tuned models or combinations of Landau levels. We show that ideal quantum geometry is directly related to the construction of fractional quantum Hall model wavefunctions.
We provide explicit analytic expressions for the flat band wavefunctions at the magic angle in terms of the CTBG wavefunctions. We also show that the Berry curvature distribution in these models can be continuously tuned  while maintaining perfect quantum geometry. Similar to the study of fractional Chern insulators in ideal $C = 1$ bands, these models pave the way for investigating exotic topological phases in higher Chern bands for which no Landau level analog is available. 
\end{abstract}

\maketitle

\emph{Introduction}--- The discovery of  superconductivity and strongly correlated phases of matter in twisted graphene-based systems  \cite{Pablo1,Pablo2,Balents2020Superconductivity} has gone hand in hand with an exploration of their unique electronic structure including topological aspects \cite{Bistritzer2011,Morell,Santos,Po2018,Po2018faithful, Khalaf_multilayer, Ahnetal,Songetal,Zhang19} emphasized by the discovery of intrinsic Chern insulating phases \cite{DavidGG,AndreaYoung,Andrei2020Graphene}. While Chern quantization is topological, band geometry controls other interaction driven phenomena, including topological mechanisms for superconductivity and fractional Chern insulators (FCIs) \cite{SkPaper,SkDMRG,Ledwith,Repellin, Lauchli21,Bergholtz20,wang2021,Xie2021FractionalCI}. Band geometry is quantified by the the Berry curvature and the Fubini-Study metric. It is of great interest to understand the interplay between band flatness, Chern number and quantum band geometry. Of particular interest are bands with higher Chern number, which have no direct Landau level analog but can be realized in twisted graphene structures \cite{Liu_2020,Cao_2020,YankowitzMonoBi1,YankowitzMonoBi2,polshyn2021topological,he2021chiralitydependent} without magnetic field, unlike Hofstadter systems \cite{Moller2015,Andrews2018,Andrews2021}.

The bands of twisted bilayer graphene, described by the Bistritzer-MacDonald (BM) model \cite{Bistritzer2011}, are greatly simplified in the ``chiral'' model introduced by Tarnopolsky {\it et al.} \cite{Tarnopolsky} where the same-sublattice moir\'e tunneling in the BM model vanishes. The chiral model has been extremely useful in understanding the physics of the system \cite{LecNotes} due to its remarkable properties: (i) perfectly flat bands at a set of magic angles, (ii) explicitly obtainable wavefunctions that are equivalent to the wavefunctions of a Dirac particle in a magnetic field \cite{Tarnopolsky, Ledwith}, (iii) wavefunctions that satisfy the ``trace condition" which relates the quantum metric to the Berry curvature; this allows the construction of Laughlin-like FCIs for short-range potentials \cite{Ledwith,wang2021}. If a band satisfies (iii) we say that it has \emph{ideal} quantum geometry. The chiral model has also served as a useful starting point in numerical studies of FCIs \cite{Bergholtz20, Repellin, Lauchli21}. Furthermore, it has inspired an improved understanding of quantum geometry of $|C| = 1$ bands in continuum models \cite{wang2021} and ideal Chern bands more broadly \cite{MeraOzawa, OzawaMera, Zhang2021,MeraDirac, MeraOzawa2,Simon2020, Varjas2021topological}.

\begin{figure}
    \centering
    \includegraphics[width = 0.4\textwidth]{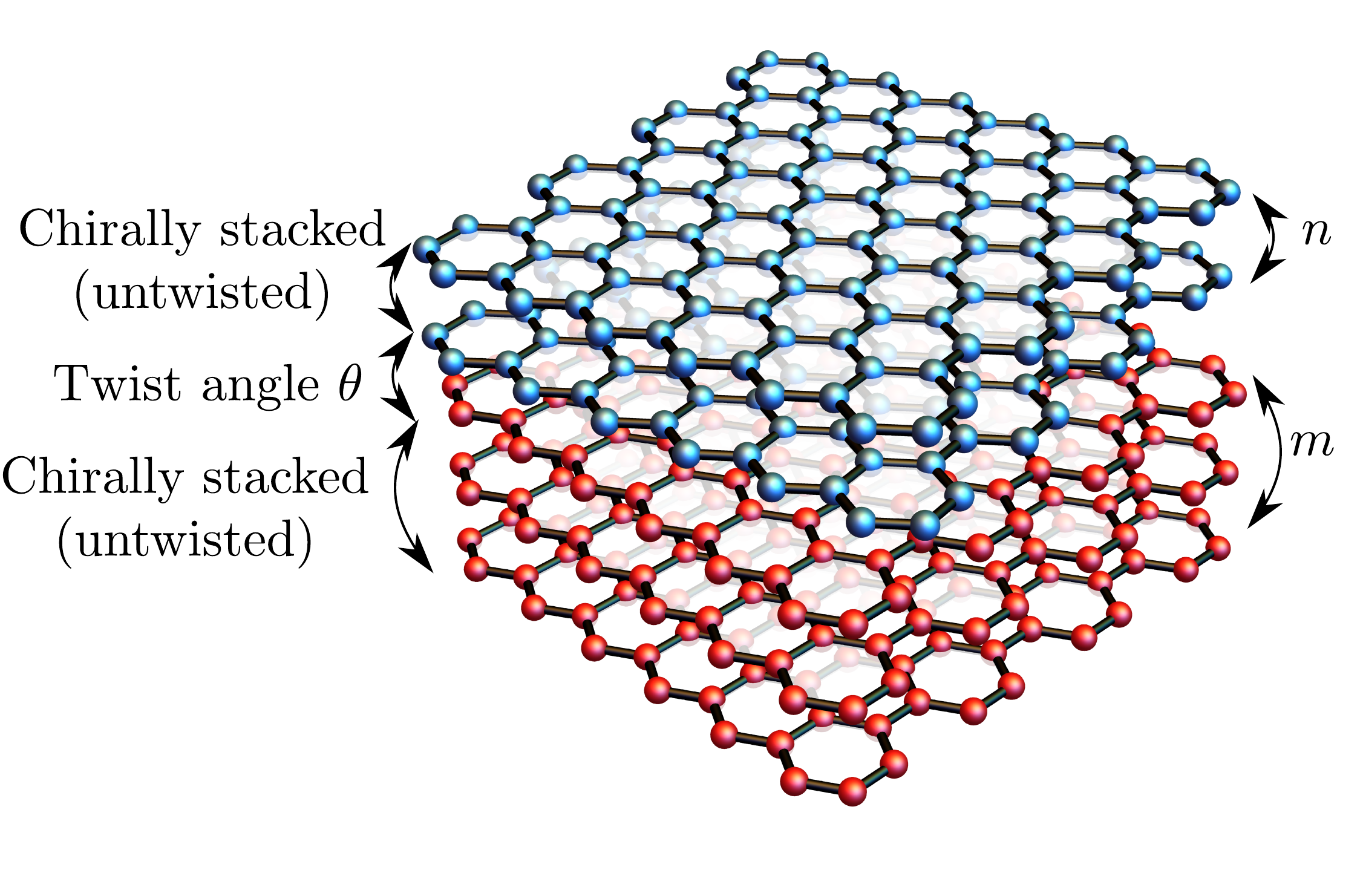}
    \caption{Schematic illustration of our multilayer setting with $n$ chirally stacked layers, e.g. AB, ABC, etc, such that each successive layers have the same Bernal stacking AB or BA, shown in blue  with a twist angle $\theta$ on top of $m$ chirally stacked layers (shown in red).}
    \label{fig:Schematic}
\end{figure}

In this Letter we describe the first ideal higher Chern bands that are not fine-tuned \cite{Claassen2015,Lee2017,Bergholtz2016} or combinations of $C=1$ bands such as the lowest Landau level (LLL)\cite{Wu2013}.
Our models are continuum models of actively explored experimental systems without external magnetic field and are not fine-tuned---their properties do not rely on a specific relationship between parameters; instead we turn off sub-leading terms in the realistic Hamiltonians. We study a class of chiral models of $n$ chirally stacked graphene layers, e.g., AB, ABC, etc, where each successive pair of layers has the same Bernal stacking AB or BA \cite{geisenhofQuantumAnomalousHall2021, shiElectronicPhaseSeparation2020, zhangSpontaneousQuantumHall2011}, twisted on top of $m$ chirally stacked graphene layers depicted in Fig. \ref{fig:Schematic} \footnote{The word chiral is being used in two distinct senses, the chiral stacking indicates the stacking pattern of graphene layers while the chiral model or limit refers to switching off same-sublattice moir\'{e} hoppings.}. Many of these structures are actively explored experimentally including twisted monolayer-bilayer graphene, $(n,m) = (2,1)$ \cite{YankowitzMonoBi1, YankowitzMonoBi2, polshyn2021topological,Morell2013ElectronicPO} and twisted double bilayer $(n,m) = (2,2)$ in both AB-AB stacking \cite{ KimTDBG, PabloTDBG, YankowitzTDBG,YahuiSenthil,LeeTDBG} and AB-BA stacking \cite{he2021chiralitydependent} 
The flat bands and their Chern numbers have been studied \cite{YahuiSenthil, DaiMultilayerChern, HaddadiTDBG, ChiralDecomposition, Graphite,TDBGJung, liuGateTunableFractionalChern2021} and anomalous Hall states have been observed \cite{Liu_2020,YankowitzMonoBi1, YankowitzMonoBi2,polshyn2021topological,he2021chiralitydependent}; it was noticed that these systems have similar magic angles to twisted bilayer graphene (TBG) \cite{YahuiSenthil,DaiMultilayerChern,Graphite,liuGateTunableFractionalChern2021}. However, the analytical nature of the wavefunctions and quantum geometry is thus far unknown.  

We show that our models realize perfectly flat bands at the \emph{same} magic angles as chiral TBG (CTBG) with ideal quantum geometry and Chern numbers $\pm n$ and $\mp m$ or $\pm 1$ and $\mp (n + m - 1)$, depending on the chirality of the stacking (e.g. AB vs BA). We also show that the ideal geometry of these models is \emph{intrinsically} $|C| > 1$; there is no decomposition into $|C|$ orthogonal ideal Chern $\pm 1$ bands. We do this by identifying a general criterion for this splittability for $C=2$ bands. Thus, our models go beyond previous idealized models of higher Chern number bands consisting of $\abs{C}$ lowest Landau levels \cite{Wu2013} while still maintaining ideal quantum geometry.

The ideal quantum geometry of these bands makes them especially suitable for realizing FCIs; we show that ideal quantum geometry enables the construction of FCI ground states through real-space holomorphicity. The ensuing realization of FCIs in higher Chern number bands would be remarkable, especially since defects in such systems, dubbed ``genons," have non-Abelian statistics \cite{Barkeshli2012,Barkeshli2013}. Unlike CTBG, these models allow for arbitrarily inhomogeneous Berry curvature which will enable future studies to pinpoint the influence of inhomogeneous Berry curvature on the stability of FCI states.

\emph{Model}--- We consider the Hamiltonian for a single graphene valley
\begin{equation}
    \H = \left(\begin{array}{cc}
    h_{n,\sigma} & T_M   \\
    T_M^\dagger & h_{m, \sigma'}
\end{array} \right)
\label{Ham}
\end{equation}
where $T_M$ is the chiral moir\'e tunneling which couples the $n$ and $n+1$ layers; it is zero except for
$
\alpha \begin{psmallmatrix}
0 & U(\br) \\
 U^*(-\br) & 0
\end{psmallmatrix} 
$ in its lower left $2 \times 2$ block.
Here, $U(\br) = \sum_{n=1}^3 e^{\frac{2 \pi i}{3} (n - 1)} e^{-i \bq_n \cdot \br}$ with $\bq_n = 2 k_D \sin \left(\frac{\theta}{2}\right) R_{\frac{2 \pi (n-1)}{3}}(0,-1)$, $k_D = \frac{4\pi}{3\sqrt{3} a_{CC}}$, and $\alpha = \frac{w}{2\hbar  v_F k_D \sin(\theta/2)}$ with $w$ the opposite sublattice moir\'{e} tunneling. All energies are measured in units of $2\hbar v_F k_D \sin \frac{\theta}{2} = \frac{w}{\alpha}$.

The Hamiltonian $h_{n,\sigma}$, with $\sigma = \pm$, takes the form of $n \times n$ block diagonal matrix (with each block having a $2 \times 2$ structure in sublattice space) given explicitly by
\begin{equation}
    h_{n,\pm} = \left(\begin{array}{cccc}
-i \bsigma \cdot \nabla & T_\pm & 0 & \dots \\
T_\pm^\dagger & -i\bsigma \cdot \nabla & T_\pm & \dots \\
0 & T_\pm^\dagger & -i\bsigma \cdot \nabla &  \dots \\
\dots & \dots & \dots & \ddots
\end{array} \right),
\label{hn}
\end{equation}
where $T_\pm = \beta \frac{\sigma_x \pm i \sigma_y}{2}$ with $\sigma_{x,y,z}$ denoting the Pauli matrices in sublattice space and $\beta=\frac{\gamma}{2\hbar  v_F k_D \sin \frac{\theta}{2}}$, where $\gamma$ is the interlayer tunneling of Bernal-stacked graphene. For realistic systems $w\approx110$ meV, $\gamma\approx360$ meV. The first magic angle occurs for $\alpha = 0.586$ at which $\beta \approx 1.9$.

The models \eqref{Ham} all have a moir\'{e} translation symmetry with lattice vectors $\ba_{1,2} = \frac{4\pi}{3}(\pm \sqrt{3}/2, 1/2)$. It is useful to define an analogue of the magnetic length $2\pi \ell^2 = A$ where $A = |\ba_1 \times \ba_2|$ is the unit cell area. The wavefunctions in layer $l$ have the Bloch periodicity \cite{Tarnopolsky, LecNotes}
$ \psi_{l,\bk}(\br + \ba_{1,2})  = e^{i (\bk - \bK_l) \cdot \ba_{1,2}} \psi_{l,\bk}(\br)$, where
$ \bK = -\bq_1$ for $l \leq n$ and $\bK' = \bq_1$ for $ l > n$ and $\bk = 0$ corresponds to the $\Gamma$ point. To incorporate these boundary conditions, we write
$
 \psi_{l,\bk}(\br) = e^{i (\bk - \bK_l) \cdot \br} u_{l,\bk}(\br),
$
 where $u_{l,\bk}(\br)$ is periodic in $\br$. 
 
 The Hamiltonian (\ref{Ham}) is purely off-diagonal in the sublattice basis, $\{\H, \sigma_z \} = 0$, and so it may be written as 
\begin{equation}
    \H = \left(\begin{array}{cc} 0 & \D^\dagger \\ \D & 0 \end{array} \right)_{\rm AB}.
\end{equation}
The ideal Chern bands will arise as zero modes of $\H$. They may be chosen to be sublattice polarized and thus zero modes of $\D, \D^\dag$.
 Note that the equation $\D \psi = 0$ is equivalent to $\tilde \D u = 0$ where $\tilde \D$ is obtained from $\D$ by replacing the $l$-th diagonal entry by $-2i \bar \partial + k - K_l$ where we use the non-bold letter $k$ to denote the complex number $k_x + i k_y$. We will use this notation for all other vectors as well. Because $\tilde \D$ only depends on $k$ and not $\ov k$ we may always choose $u_k$ that are zero modes of $\tilde{D}$ to be holomorphic functions of $k$ as well \cite{Ledwith, LecNotes}.
 
 \textit{Band Geometry ---}
 Here we say that the quantum geometry of a band is \emph{ideal} if the band satisfies the trace condition. Without loss of generality we take $C>0$; complex conjugation may be applied to obtain analogous statements for $C<0$. The trace condition is the saturation of the inequality
 \begin{equation}
     \tr g(\bk) \geq \Omega(\bk)
    \label{trcond}
 \end{equation}
 where $g$ is the Fubini-Study metric and $\Omega$ is the Berry curvature. The trace condition is necessary for reproducing lowest Landau level physics \cite{Claassen2015,Roy2014,Parameswaran2013,Ledwith,LecNotes,wang2021} and it holds if and only if the wavefunctions $u_\bk$ can be chosen as holomorphic functions of $k_x + i k_y$ \cite{wang2021,MeraOzawa}. If the trace condition holds and the Berry curvature is homogeneous then the density operators satisfy the Girvin-Macdonald-Platzmann algebra \cite{gmp1986} of the LLL \cite{Roy2014,Parameswaran2013}.
 
 We now show that the trace condition enables the construction of model fractional quantum Hall wavefunctions. The trace condition implies that $z\psi = Pz\psi$ for $z=x+iy$ and $P = \sum_\bk \ket{\psi_\bk} \bra{\psi_\bk}$ the projector onto the band of interest \cite{Roy2014,Varjas2021topological}.  Through iteration and power series we have $f(z)\psi = Pf(z)\psi$; multiplication by holomorphic functions does not take the wavefunction out of the band of interest. For many-body wavefunctions, we may then attach factors such as $\prod_{i<j} (z_i - z_j)^n$ without involving remote bands. The ability to attach such factors implies fractional quantum Hall ground states in flat bands with short range interactions \cite{Haldane1983,TrugmanKivelson,Ledwith,LecNotes}. So far, this conclusion has only been derived for the case of ideal $C=1$ bands where the wave functions are related to LLL wavefunctions \cite{wang2021}. Our models allow for the extension of these ideas to higher Chern bands.
 
 While the trace condition is sufficient to guarantee a FCI ground state in the limit of short ranged interactions, this limit may be difficult to reach when the Berry curvature is inhomogeneous. In the extreme limit of soleinoidal Berry curvature the band can energetically resemble a band of different Chern number \cite{SoftModes}; this is easiest to see in a finite-size system where a concentrated Berry curvature is invisible to the momentum space grid. 
 For the models we study, the Berry curvature may be tuned to be as inhomogeneous as one wishes, enabling future works to isolate the effect of inhomogenous Berry curvature while preserving the trace condition.

\emph{The Foundation: A Review of CTBG}---
The foundation of the solution to the general multilayer models is $n=m=1$, or CTBG. Here we review the wavefunctions of CTBG on the $A$ sublattice \cite{Tarnopolsky,Ledwith} that form a band of zero modes of the operator
\begin{equation}
    \D = \left(\begin{array}{cc} 
 -2i \bar \partial & \alpha U(\br) \\
 \alpha U(-\br) & -2i \bar \partial
\end{array}\right).
\label{Dtbg}
\end{equation}
The $B$ sublattice wavefunctions are related by $C_2 \T: \psi(\br) \mapsto \ov{\psi(-\br)}$. 

We focus on writing the periodic wavefunctions $u_k$ in terms of the (modified \cite{Haldane2018}) Weierstrass sigma function \cite{wang2020chiral,wang2021} $\sigma(z) = \sigma(z|a_1, a_2)$. It satisfies
  $\sigma(z) = -\sigma(-z)$ and $ \sigma(z+a_{1,2}) = - \exp\left(\frac{1}{2\ell^2}a_{1,2}^*(z+\frac{a_{1,2}}{2} ) \right)\sigma(z)$ which together imply $\sigma(a) = 0$ for all lattice vectors $\ba$.

The function $\phi_k(\br) = e^{-\frac{i}{2}z^* k }\sigma(z + i\ell^2k)$ satisfies
  $\phi_{\bk + \bb_i}(\br) = e^{-i \bb_i \cdot \br } e^{i \theta_{k ,b_i}} \phi_k(\br)$ with $\theta_{k,b} = \pi - \frac{1}{2}i\ell^2 b^*(k-b/2) $
and is a building block for all the models in this paper. 

The chiral TBG periodic wavefunction may be written as
\begin{equation}
  u_k(\br) = \phi_\bk(\br) \frac{u_\Gamma(\br)}{\sigma(z)} = \phi_\bk(\br) e^{-K(\br)} \bn(\br).
  \label{chiralTBG}
\end{equation}
Without the normalized layer spinor $\bn(\br)$, this wavefunction is that of a Dirac particle moving in an inhomogeneous magnetic field $B(\br) = \nabla^2 K(\br)$ with one flux quantum of flux per unit cell \cite{Ledwith}. The normalized layer spinor drops out of all Bloch overlaps and therefore does not influence the quantum geometry of the system or the interacting physics for density-density interactions.

Throughout this paper we consider wavefunctions that are smooth in $k$ but not periodic; one may always choose such a gauge. The Chern number may then be computed by taking the line integral of the Berry connection around the Brillouin zone and using the $\bk$-space boundary conditions. One obtains \cite{LecNotes,wang2021}
\begin{equation}
  C = \frac{1}{2\pi} \Re (\theta_{k+b_1,b_2} - \theta_{k,b_2} + \theta_{k,b_1} - \theta_{k+b_2,b_1} ).
  \label{ChernnumberBCs}
\end{equation}
For CTBG we see that $C=1$.

\emph{Simple example: chiral twisted monolayer-bilayer graphene}--- We now show that the Hamiltonian (\ref{Ham}) has perfectly flat bands at the same set of magic angles as chiral TBG. We start with $n = 2$, $m = 1$ and $\sigma = +$, chiral twisted monolayer-bilayer. The zero mode operator is 
\begin{equation}
    \D(\br) =
\left(\begin{array}{ccc} 
-2i \bar \partial & \beta & 0  \\
0 & -2i \bar \partial & \alpha U(\br) \\
0 & \alpha U(-\br) & -2i \bar \partial
\end{array}\right).
\label{HMonoBi}
\end{equation}

Let us start with sublattice A. The equations from the second and third rows of $\D \psi = 0$ are identical to those of CTBG \eqref{Dtbg}.
Thus
, we can write a solution to $\D \psi = 0$ with $\psi = (\psi_1, \psi_2, \psi_3)$ as follows:
\begin{equation}
    \psi_{2,\bk} = \lambda_\bk \psi^{\rm TBG}_{1,\bk}, \quad  \psi_{3,\bk} = \lambda_\bk \psi^{\rm TBG}_{2,\bk}, \quad  2i \bar \partial \psi_{1,\bk} = \beta \lambda_\bk \psi^{\rm TBG}_{1,\bk}
    \label{PsiTBG}
\end{equation}
where $\lambda_\bk$ is a $\bk$-dependent constant to be determined soon. 

In Fourier space, $u_\bk(\br) = \sum_\bG e^{i\bG \cdot \br} u_\bk(\bG)$, the last equation may be solved:
 \begin{gather}
  u_{1,\bk}(\bG) =- \frac{\beta \lambda_\bk}{k - K + G} u^{\rm TBG}_{1,\bk}(\bG), \\
  u_{2,\bk}(\bG) = \lambda_\bk u^{\rm TBG}_{1,\bk}(\bG), \quad u_{3,\bk}(\bG) = \lambda_\bk u^{\rm TBG}_{2,\bk}(\bG).
  \label{uk}
 \end{gather}
 As discussed above, because \eqref{uk} gives a band of zero modes of the operator $\tilde{\mathcal{D}}$ we may choose $u_\bk$ and $\lambda_{\bk}$ to only depend on $k$ such that the band satisfies \eqref{trcond}.
 
 We now describe $\lambda_k$ and obtain the Chern number. In order for \eqref{uk} to be normalizable\footnote{Note that $u_\bk$ and $u^{\rm TBG}_\bk$ are in general not normalized if we choose them to be holomorphic functions of $k$, but they should be normalizable}, we need $\lambda_k$ to have a single zero at $k = K-G$ and no others. 
 This fixes the form $\lambda_k = \phi_{k-K}(0)$ up to gauge transformations. 
 Since the phase of $\lambda_\bk$ winds by $2\pi$ around the BZ, multiplication by $\lambda_\bk$ increases the Chern number by 1 compared to the TBG bands. Thus, $u_k$ has Chern number 2. We may also compute the Chern number from the boundary condition method \eqref{ChernnumberBCs} and obtain $2$ as well.
 
 Let us now consider the B sublattice. 
 Writing the operator $\D^\dagger$ explicitly
 \begin{equation}
     \D^\dagger =
\left(\begin{array}{ccc} 
-2i \partial & 0 & 0  \\
\beta & -2i \partial & \alpha U^*(-\br) \\
0 & \alpha U^*(\br) & -2i \partial
\end{array}\right)
 \end{equation}
 There is a zero energy state given by $\psi_B = (0, \psi^{\rm TBG}_{B,1}, \psi^{\rm TBG}_{B,2})$ where $\psi^{\rm TBG}_B(\br) = [\psi^{\rm TBG}_A(-\br)]^*$. Thus, the B sublattice wavefunctions are identical to the B sublattice wavefunctions of CTBG and have Chern number $-1$. Our result of two flat bands per valley with Chern numbers $\pm 2$ and $\mp 1$ is compatible with realistic twisted monolayer-bilayer graphene \cite{polshyn2021topological}.
 
\begin{figure}
     \centering
     \includegraphics[width = 0.49\textwidth]{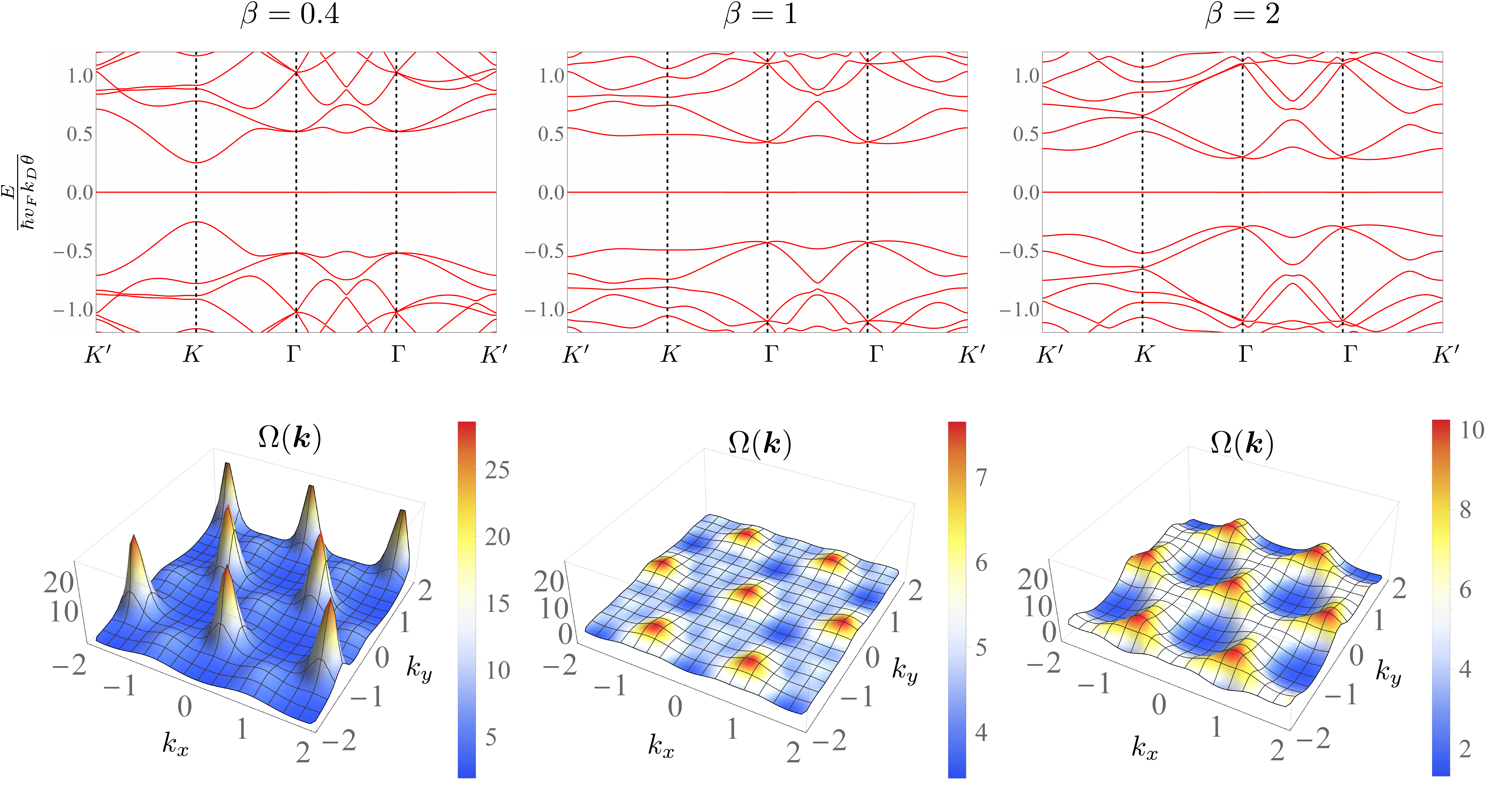}
     \caption{ {\bf Band structure and Berry curvature for chiral twisted monolayer-bilayer graphene}: Band structure for the Hamiltonian (\ref{HMonoBi}) at the first magic angle for $\beta = 0.4, 2$ (top panel) and Berry curvature distribution for the $C = 2$ band for $\beta = 0.4, 1, 2$ (bottom panel). We see that for small $\beta$, the Berry curvature is strongly peaked at the $K$ point and by increasing $\beta$, it gets more uniform with the peak moving to the $\Gamma$ point. It is easy to check numerically that the trace condition $\Tr g(\bk) = |\Omega(\bk)|$ is always satisfied.}
     \label{fig:MonoBi}
 \end{figure}

 The previous analysis implies that the Hamiltonian (\ref{HMonoBi}) has the same magic angles as TBG: remarkably the angles are independent of the interlayer coupling $\beta$. This is illustrated in Fig.~\ref{fig:MonoBi}a-c, which shows the band structure at the first magic angle for different values of $\beta$. Although the bands remain flat, the overall band structure and flat band wavefunctions depend on the parameter $\beta$. 
 As $\beta \to 0$ the band gap closes and Berry curvature diverges as shown in Fig.~\ref{fig:MonoBi}e-g. and the system decouples into $C=\pm1$ CTBG and graphene.

\emph{No decomposition into two ideal $C=1$ bands}--- We now show that upon doubling the unit cell with translation breaking wavevector $\bQ$, an ideal $C=2$ band can be decomposed into two ideal orthogonal $C=1$ bands if and only if the Berry curvature satisfies $\Omega(\bk) = \Omega(\bk+\bQ)$. Chiral twisted monolayer-bilayer graphene does not satisfy this condition.

We assume, and soon contradict, that the wavefunctions of the two $C=1$ bands $\ket{\tilde{u}_{k \zeta}}$ for $\zeta=1,2$ are holomorphic in $k$ and therefore may be written as
     $\ket{\tilde{u}_{\zeta k}} = \alpha_{\zeta k} \ket{u_{k}} + \beta_{\zeta k} e^{i \bQ \cdot \br} \ket{u_{k+Q}}$
for some holomorphic $\alpha_{\zeta k}$ and $\beta_{\zeta k}$. 
The orthogonality $\bra{\tilde{u}_{k2}}\ket{\tilde{u}_{k1}} = 0$  implies $-\frac{\beta_{k1}\ov{\beta_{k2}}}{\alpha_{k1}\ov{\alpha_{k2}}} = \frac{\norm{u_k}^2}{\norm{u_{k+Q}}^2}$, where we used $\bra{u_k} e^{i \bQ \cdot \br} \ket{u_{k+Q}} = 0$ and we work in a patch of $k$ space away from zeros of $\alpha_{\zeta}$. The right hand side is positive; thus we have $\frac{\beta_{2k}}{\alpha_{2k}} = -c \frac{\beta_{1k}}{\alpha_{1k}}$ for a real positive $c$. Let us define $e^{w_k} = \beta_{1k}/\alpha_{1k}$; then we have $\Re w_k = \log \frac{c \norm{u_k}}{\norm{u_{k+Q}}}$. Real parts of holomorphic functions are harmonic (have zero laplacian), and a holomorphic function may be reconstructed from a harmonic real part. Therefore the decomposition is possible and unique if and only if $\nabla^2 \log \frac{\norm{u_k}}{\norm{u_{k+Q}}}=\Omega(\bk) - \Omega(\bk + \bQ)= 0$ for all $\bk$; here we used $\Omega(\bk) = \nabla^2 \log \norm{u_k}$ \cite{wang2021,MeraOzawa2}.
  
 \emph{General case}--- We now generalize to arbitrary $n$, $m$, $\sigma$ and $\sigma'$. It is sufficient to focus on sublattice A and $(\sigma,\sigma') = (+,+), (+,-)$ and $(-,+)$. Any sublattice B band may be mapped to a sublattice A band using $C_{2z} \T$, under which the valley is kept invariant, sublattices are exchanged, Chern number switches sign and layers are kept invariant: $(n,m,\sigma,\sigma')\mapsto(n,m,-\sigma,-\sigma')$. Next, we may map $(-,-)\to(+,+)$ stacking by $C_{2y} \T$ under which the valley, sublattice, and Chern number are kept invariant while $l \mapsto n + m - l + 1$ which switches chirality: $(n,m,\sigma,\sigma')\mapsto(m,n,-\sigma',-\sigma)$.

 We first consider $(\sigma, \sigma') = (+, -)$. As before $\psi_n = \lambda_\bk \psi_1^{\rm TBG}$ and $\psi_{n+1} = \lambda_\bk \psi_2^{\rm TBG}$. The remaining components are given by solving the equations $2i \bar \partial \psi_l = \beta \psi_{l+1}$ for $l < n$ and $2i \bar \partial \psi_l = \beta \psi_{l-1}$ for $l > n + 1$. These equations have the solution
  \begin{gather}
     u_{l,\bk}(\bG) = \lambda_\bk \left( \frac{-\beta}{k - K + G}\right)^{n-l} u_{1,\bk}^{\rm TBG}(\bG), \quad l \leq n \label{ul1}\\
     u_{l,\bk}(\bG) = \lambda_\bk \left( \frac{-\beta}{k - K' + G}\right)^{m-(l-n)} u_{2,\bk}^{\rm TBG}(\bG), \quad l > n \label{ul2}
 \end{gather}
 This yields a normalizable wavefunction if and only if $\lambda_k$ has a zero of order $n - 1$ whenever $\bk = \bK-\bG$, a zero of order $m - 1$ whenever $\bk = \bK'-\bG$, and no others which gives a total Chern number of $n + m - 1$. 
We have
     $\lambda_k  = \phi_{k-K}(0)^{n-1} \phi_{k-K'}(0)^{m-1}$. For $l < n$, the wavefunction $u_{l k}(\br)$ contains the factor $\phi_{k-K}(0)^{l-1}\phi_{k-K'}(0)^{m-1}$, as well as $m-1$ zeros at $k=K'$. Analogous considerations apply to the case $l>n$.

For $(\sigma,\sigma')=(+,+)$ we may set $u_{kl} = 0$ for $l>n+1$. The $l\leq n$ wavefunctions are then the same as the $(\sigma,\sigma')=(+,-)$ case with $m=1$. Finally, for $(\sigma,\sigma') = (-,+)$ we may set $\psi_{l<n}$ and $\psi_{l>n+1}$ to zero and recover chiral TBG wavefunctions. A summary of the results is provided in Table \ref{tab:Chern}. Note that the wavefunctions $u_k$ are always analytic in $k$ which means that the bands always have ideal quantum geometry.
 
 
 \begin{table}[]
     \centering
     \begin{tabular}{c|c|c}
     \hline \hline
        $(\sigma, \sigma')$ & Chern A & Chern B  \\
        \hline
        $(+,+)$ & $n$ & $-m$\\
        $(-,+)$ & 1 & $-(n + m - 1)$ \\
        $(+,-)$ & $n + m - 1$ & $-1$\\
        $(-,-)$ & $m$ & $-n$ \\
        \hline \hline
     \end{tabular}
     \caption{Chern numbers for the A and B sublattice bands for a configuration of $n$-layers twisted on top of $m$-layers.  }
     \label{tab:Chern}
 \end{table}

\begin{figure}
     \centering
     \includegraphics[width = 0.36\textwidth]{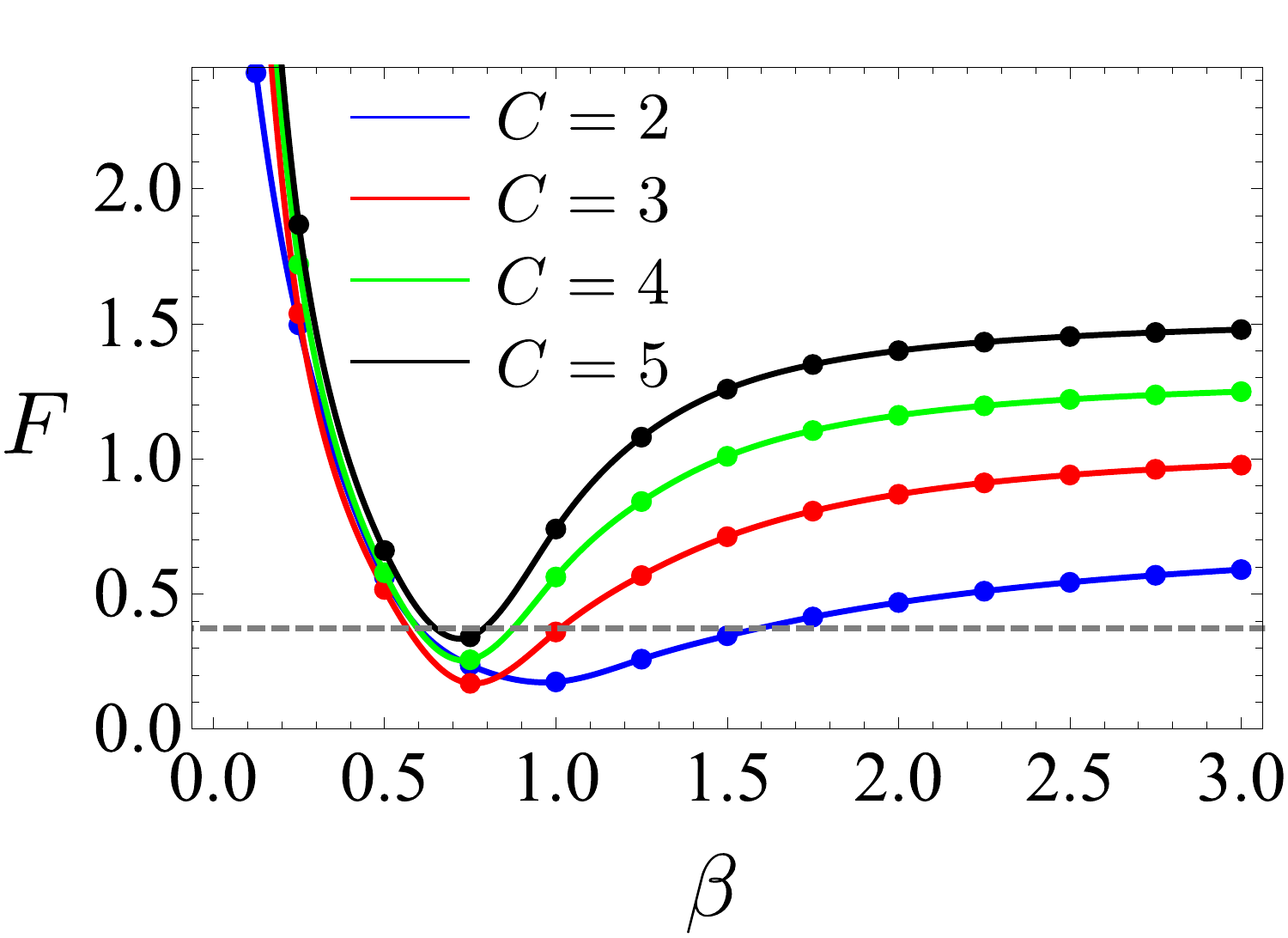}
     \caption{{\bf Berry deviations:} Plot of the Berry inhomogeneity $F$ defined in Eq.~\ref{dF} as a function of $\beta$ for $C = 2,3,4,5$ obtained from models with $n = 2,3,4,5$ and $m=1$. The dashed line indicates the corresponding value of $C = 1$ CTBG bands.}
     \label{fig:Berry}
 \end{figure}

 \emph{Berry curvature variations}--- The models introduced here provide a realization of Chern bands that satisfy the trace condition with arbitrary Chern $C$ where the Berry curvature is continuously tunable and arbitrarily inhomogeneous. This is illustrated in Fig.~\ref{fig:Berry} by plotting the Berry deviation
\begin{equation}
    F = \left(\int \frac{d^2\bk}{A_{\rm BZ}} \left[\frac{A_{\rm BZ}\Omega(\bk)}{2\pi C} - 1\right]^2\right)^{1/2}
    \label{dF}
\end{equation}
 for bands with $C = 2,3,4,5$ as a function of the Bernal-stack coupling parameter $\beta$. As we can see, $F$ diverges in the decoupled $\beta \rightarrow 0$ limit. The minimal value of F occurs around $\beta \approx 0.75 - 1$, and for the realistic $\beta \approx 1.9$ the inhomogeneity is not very large.
 
 
\emph{Conclusion} Here we have shown that a family of chirally twisted graphene structures can, in a particular limit, realize flat and ideal Chern bands with arbitrary Chern numbers. This setup has a new tuning parameter which strongly affects Berry curvature distribution while keeping the ideal quantum band geometry intact despite having the same magic angle as twisted bilayer graphene. Although the ideal limit discussed here is not perfectly realized in actual materials, additional terms like a displacement field may help access this limit in realistic systems. Independent of their realizability, our models are a promising starting point for exploring exotic topological phases at fractional filling of ideal flat higher Chern bands whose interaction physics is poorly understood due to the lack of a Landau level analog.

\emph{Acknowledgements} We thank Dan Parker and Tomohiro Soejima for several insightful discussions and collaborations on related projects. AV was supported by a Simons Investigator award and by the Simons Collaboration on Ultra-Quantum Matter, which is a grant from the Simons Foundation (651440,  AV). P.J.L. was supported by the Department of Defense
(DoD) through the National Defense Science
and Engineering Graduate Fellowship (NDSEG)
Program.

\emph{Note Added} During the completion of this work, Ref \cite{WangMultilayer} appeared, which overlaps with the results reported here.  

\bibliographystyle{unsrt}
\bibliography{refs}

\end{document}